# Novel photo-conducting state and its perturbation by electrostatic fields in oxide-based two-dimensional electron gas

A. Rastogi[1], J. J. Pulikkotil[2], S. Auluck[2], Z. Hossain[1] and R. C. Budhani[1,2]*
[1]*Condensed Matter, Low Dimensional Systems Laboratory,*
*Department of Physics, Indian Institute of Technology Kanpur, 208016, India*
[2]*National Physical Laboratory, Council of Scientific and Industrial Research,*
*Dr. K. S. Krishnan Marg, New Delhi 110012, India*
(Dated: August 4, 2012)

The two-carrier transport model as proposed for the two-dimensional electron gas formed at the interfaces of oxide heterostructures is investigated by means of a combined perturbation of near ultra-violet radiation and electrostatic field, applied both separately and simultaneously. Comparison of the photo-response of the prototype systems such as the band insulator $LaAlO_3$ and Mott insulator $LaTiO_3$ films on $TiO_2$ terminated $SrTiO_3$ show remarkably similarities. Two types of non-equilibrium carriers are generated in each system, having a signature of a particular type of perturbation characterized by distinctly different relaxation process. While, the photo-conducting state diminishes in a stretched exponential manner, with a temperature dependent activation energy varying from few tens of meV to $\approx 1$ to 2 meV on lowering the temperature, and a relaxation time of several hours, the recovery from electrostatic gating occurs in milli-seconds time scale. An attempt is also made to explain the experimental observations using the *ab-initio* density functional calculations. The calculations show that the electronic transitions associated with near ultra-violet radiation emerge from bands located at $\simeq 2$ eV above and below the Fermi energy, which are the Ti 3d states of the $SrTiO_3$ substrate, and that from the $AlO_2$ ($TiO_2$) layers of the $LaAlO_3$ ($LaTiO_3$) films, respectively. The slow decay of the photo-current to the unperturbed state is explained in terms of the closely spaced Ti $3d_{xy}$ states in the lower conduction band, which are manifested as flat bands (or localized states) in the band structure. Such localization leads to increased carrier life-times, through the energy-time relationship of the uncertainty principle.

## I. INTRODUCTION

Polarization discontinuity at the interface of oxide heterostructures, such as that of $LaAlO_3$ and $LaTiO_3$ on $TiO_2$ terminated $SrTiO_3$, induces large in-built electric field [1, 2]. This thermodynamic instability is nullified by one or more mechanism(s) such as atomic relaxation[3], electronic re-construction [4, 5], inter-layer cation mixing [6, 7] and/or creation of vacancies [8, 9], thereby resulting in the formation of two-dimensional electron gas (2DEG) [1, 10–12]. Properties of such 2DEG state have been extensively studied in the recent past, owing to its display of magnetism, superconductivity, and fractional quantum Hall effects [8, 9, 13–18]. Thus, the subject of addressing the origin and nature of these charge carriers has captured much attention. Brinkman *et. al.* propose that the 2DEG comprises of two different types of charge carriers, which fundamentally clears the co-existence of two order parameters, namely superconductivity and magnetism [15], with the latter being associated with the localized electrons of the Ti $3d$ orbitals and superconductivity being mediated via the itinerant electrons. A two-band model has been proposed to describe the dependence of sheet resistance and Hall coefficient under the influence of magnetic field [19]. The two-carrier model is also supported by means of density functional calculations [20, 21] and high field Hall measurements[25].There are also evidences, both for and against, multi-band model with two electron populations to describe the transport at the oxide interfaces [22–24]. Beyond, it has also been observed that the properties exhibited by the 2DEG at the interfaces can be altered by an electrostatic potential applied in a field effect geometry [17, 27–30].

From the electronic transport perspectives, the films of $LaAlO_3$ (LAO) and $LaTiO_3$ (LTO) grown on $TiO_2$ terminated $SrTiO_3$ (STO) substrates with film thickness of 20Å ($\simeq 4$ unit-cells ) and above, shows a sheet carrier density of $\sim 10^{13}$ cm$^{-2}$ [8, 9, 15–17]. These values are significantly lower as compared to that reported in the original works of Ohtomo and Hwang [1], which is $\sim 10^{17}$ cm$^{-2}$ . The discrepancy seems to arise from the synthesis techniques involving growth temperature and oxygen pressure during the growth of the films. While the high carrier densities are associated with the oxygen vacancies which source the extra electrons, the state of reduced carrier appears as an intrinsic property of the atomically abrupt interfaces. For such low carrier deficient systems, time-resolved photoluminescence experiments [30, 31] have shown a pronounced photoconductivity activated by photon of wavelength of $\lambda \leq$ 400 nm. The ultra-violet (UV) sensitive enhancement in conductivity has many characteristics that are similar to those seen in III-IV compound semiconductor interfaces such as GaAlAs/GaAs [32, 33], InGaP/GaAs

---

*Electronic address: rcb@nplindia.org

[34]and GaAlN/GaN [35]. Besides, a remarkable feature of the photo-conductivity in these oxide heterostructures is their inordinately slow recovery to the unperturbed state on extinguishing the photo-illumination.

Due to the inherent ability of transition metal ions to display multi-valent states, defects in the form of O vacancies would lead to electronic reconstruction in their neighborhood to preserve the charge neutrality. If such a scenario leads to vacancy derived localized states, they would serve as centers to optical excitations. If in case the centers are singly charged they act as recombination centers for minority carriers, while those which are doubly (or multiply) charged become traps [36]. However, in these oxide heterostructures it is observed that trapping centers exist even in the case of abrupt interfaces [30], thereby suggesting a mechanism different from that of the defect mediated ones. *Ab-initio* calculations carried out for the fully relaxed abrupt interfaces reveal a series of closely spaced Ti $3d_{xy}$ states in the lower conduction band. These states, which are derived from the STO substrate, are highly localized and therefore tend to act as traps due to their increased carrier life-times. The above consensus is derived from the energy-time relationship of the uncertainty principle. Beyond, our finding that the change in resistance due to electric field being independent of the recovery of the photo-conducting state confirms that the radiation and electric fields act on two different sources of carriers, and thus are partly decoupled in these heterostructures.

## II. EXPERIMENTAL AND COMPUTATIONAL DETAILS

### A. Experiments

Ultra thin films of LAO and LTO were deposited on (001) cut STO single crystal substrate by ablating well-sintered targets of these materials with nanosecond pulses of an excimer laser (KrF, $\lambda$= 248 nm). It has been shown that properties of such films depend strongly on growth conditions and a metallic interface is realized in films deposited on $TiO_2$ terminated surfaces. We achieved such termination by etching the substrate in buffered HF solution followed by cleaning in de-ionized water and alcohol for 20 minutes. The substrate was annealed at 800$^o$C for one hour in $7.4 \times 10^{-2}$ mbar of oxygen to get a defects free surface. The LTO and LAO films of thickness 8 nm were deposited at 700, 750, and 800$^o$C in $1 \times 10^{-4}$ mbar of $O_2$ with a slow rate ($\sim$ 0.12Å/s) to realize a layer-by-layer growth. The films were cooled under the same pressure at the rate $10^o$/min. Two sets of samples are used for the measurements; in one case 2.5 $\times$ 5 mm$^2$ slabs were prepared for standard four probe measurements of resistance and photo-conductivity by depositing Ag/Cr contact pads through shadow mask such that the separation between voltage pads is $\simeq$ 700 $\mu m$. For field effect measurements, we have used 5 $\times$ 5 mm$^2$ samples. Here the source and drain were deposited to give an effective channel area of 200 $\times$ 100 $\mu m^2$, whereas the gate electrode was coated on the backside of the 0.5 mm thick STO substrate. The integrated UV intensity $\simeq 8\mu W cm^{-2}$ of a quartz halogen lamp and He-Cd laser lines of 441 and 325 nm were used for photo-illumination while the sample was mounted in a close cycle helium cryostat.

### B. Theoretical calculations

On the theoretical front, the calculations were carried out using the full-potential linearized augmented plane wave (FP-LAPW) method, as implemented in the Wien2k suite of programs [37]. The LAO/STO heterostructures were modeled by stacking 5 unit-cells of STO (representing substrate) with four unit-cells of LAO (representing film), with cell dimensions being that of STO (a = 3.905 Å). The structural parameters were fully relaxed using force optimization technique. The exchange-correlation potential was described in the generalized gradient approximation (GGA). Convergence in the basis set was achieved with $RK_{max}$=6.0, with R being the smallest LAPW sphere radius and $K_{max}$ being the interstitial plane-wave cut-off. The sphere radius, in Bohr units, for Sr/La, Ti/Al and O were chosen as 2.2, 1.9 and 1.6, respectively. Convergence of the Brillouin zone (BZ) sampling is attained for Monkhorst–Pack $k$-mesh of 90 $k$- points in the irreducible region of the BZ.

## III. RESULTS AND DISCUSSION

### A. Electrical transport

We first compare the sheet resistance ($\Omega/\square$) of LAO and LTO films grown at 700, 750, and 800 $^o$C on the $TiO_2$ terminated STO substrate. The data shown in Fig. 1(a & b) are plotted on the logarithmic scale to emphasize the temperature dependence down to 20 K. The temperature dependence of resistance measured between 300 K and 20 K shows metallic behavior for all films. The metallicity becomes robust for the films grown at higher temperatures. It is also interesting to note that while the sheet resistance of LAO/STO tends to saturate or even grows marginally below $\approx$ 50 K, the resistance of LTO/STO films continues to drop down to 20 K. The temperature dependence of the resistance of LTO/STO systems can be expressed as R = $R_0$+ $AT^2$, whereas for LAO/STO, a BlnT term needs to be added where B is a negative coefficient. As evident from the inset of Fig. 1(a), the room temperature resistance is lesser for LTO/STO, in comparison to the LAO/STO systems synthesized at 800$^o$C. With structural symmetry and lattice parameters of both LaTiO$_3$ and LaAlO$_3$ being similar and that they have $\approx$2% of lattice mismatch with SrTiO$_3$, the difference in the R(T) may be then accounted to the nature of electronic interactions in the films. We note that while LaTiO$_3$ is a Mott insulator, LaAlO$_3$ is a band insulator. Therefore, it may be inferred that the Ti 3d electrons which are strongly correlated in LaTiO$_3$ play an important role in the low temperature transport properties. It is also interesting to note that these films become superconducting, at lower temperatures (T < 300 mK) [18].

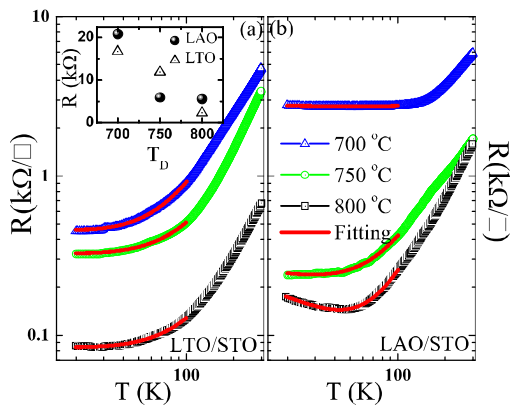

Figure 1: (Color Online) The figure shows the temperature dependence of the sheet resistance down to 20 K for 8 nm thick LTO/STO (panel a) and for LAO/STO (panel b) films deposited at 700, 750 and 800$^o$C on STO (001) substrate. Solid lines (red lines) are fitted to the expression R = $R_0$+A$T^2$+ BlnT, where $R_0$, A, and B are constants, over the temperature range 20 to 100 K. For LTO/STO, B=0. A comparison of the room temperature sheet resistance of LTO and LAO films as a function of growth temperature is shown in the inset of panel (a).

### B. Photo-conductivity

The resistance of LTO and LAO films shows nearly similar photo-conductive behavior on exposure to the light from a quartz halogen lamp and also shows a similar decay behavior after removing the exposure. Using He-Cd laser of wavelengths 441 and 325 nm, we have further established that the photo-conductivity emanates only from the UV component of the halogen lamp. The inset of Fig. 2(a) shows the temporal dependence of the change in resistance $\Delta R/R_D$, where $\Delta R$ = $R_D$ - R(t) and $R_D$ is the resistance just before the exposure (point A in the figure), R(t) is the resistance at time t. The light of UV intensity ≈ 8 $\mu W/cm^{-2}$ is turned off at point B and isothermal recovery of resistance is monitored as a function of time. The main panel of Fig. 2(a) shows the isothermal decay of the photo-conducting state of the LTO film deposited at 800$^o$C. A similar set of data taken at different temperatures for the LAO film is shown in Fig. 2(b). The absolute value of $\Delta R/R_D$ just before ex-tinguishing the exposure (corresponding to the point B in 2 (a) inset) at 300 and 20 K for films grown at three different temperatures is plotted in the inset of Fig. 2(b). We note a distinctly larger change in the resistance of LTO films on photo exposure.

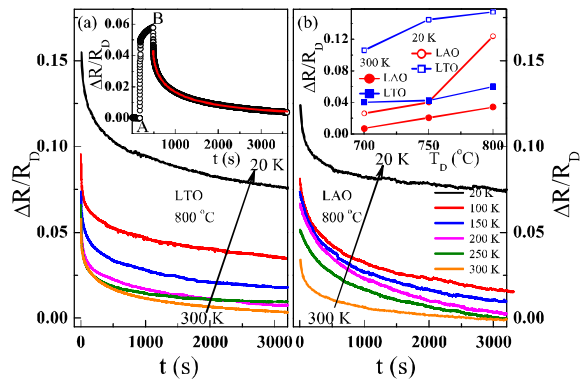

Figure 2: (Color Online) The temporal evolution of the normalized resistance $\Delta R/R_D$, as defined in text, after turning off the light for 800$^o$ C deposited LTO/STO and LAO/STO films at different temperature is shown in (a) and (b). Inset of (a) defines the instant at which the UV exposure was initiated and extinguished. At point 'A' the exposure is switched on, at 'B' it is turned off and the resistance is monitored as a function of time. The main panel shows the normalized resistance after point 'B'. The inset of (b) shows the growth temperature dependence of the maximum relative change in resistance for both the heterostructures at 20 and 300 K.

In Fig. 3 we compare the decay of photo-conducting state in the films deposited at 700$^o$C. From these data and also from the inset of Fig. 2(b) it is clear the photo-response is higher in LTO films irrespective of the growth temperature. This observation suggests that the intrinsic property of LTO to oxidize takes away oxygen from the substrate rendering it photosensitive.

### C. The Kohlrausch model

The time dependent normalized resistance was fitted to the stretched exponential given by Kohlrausch [38] ; $\frac{\Delta R(t)}{R_D} \sim \exp\left[-\left(\frac{t}{\tau}\right)^\beta\right]$, where $\tau$ is the relaxation time constant and $\beta$ the decay exponent which takes values between 0 to 1 [38–40]. The $\beta$ = 1 situation describes the Debye relaxation process where a single activation energy is involved. A semi logarithmic plot of time constant $\tau$ as a function of (1/T) for the films grown at three different temperatures is shown in Fig. 4 along with the behavior of $\beta$. We note that the system recovers much faster at the higher temperatures of measurement. The exponent $\beta$ also approaches a Debye like behavior ($\beta$= 1) from very small value of ≃ 0.2 at 20 K. The ln($\tau$) vs (1/T) plot sug-gests a thermally activated behavior $\tau \sim \tau_0 \exp\left[-\frac{\Delta U}{k_B T}\right]$ where $\Delta U$ is a function of temperature and $k_B$ the Boltzmann constant. A simplified picture of the recovery can be presented based on two activation energies shown in Fig. 4(e) and (f). The energy changes from $\approx 1$ meV to 11 meV as we go from the low temperature to high temperature regime for LTO system, while ranges from $\approx 1$ meV to 20 meV for LAO system. This order of magnitude difference in U(T) at low and high temperature suggests to a different dynamics of the migration of photo-excited carriers.

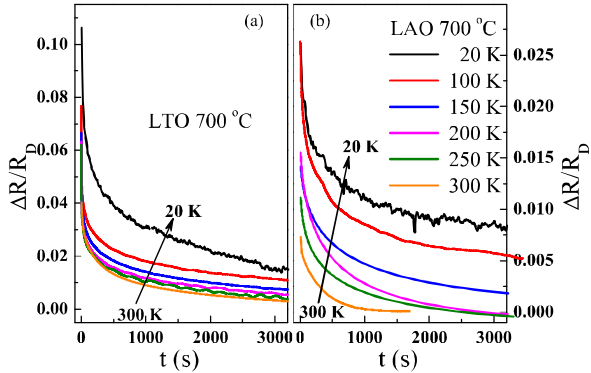

Figure 3: (Color Online) The time evolution of the normalized resistance at several temperatures is shown for (a) LTO/STO, and (b) LAO/STO deposited at 700° C. The change in resistance in case of LTO films is larger in comparison to that for the LAO. For both the films deposited at 700° C the magnitude of the photo-response is comparatively less to that seen in films deposited at 800° C.

The two different regions in the Arrhenius plot (Fig. 4e and 4f) can be explained within the frame work of the large lattice relaxation model [41] proposed earlier to understand the recovery from the photo-conductive state in III-V compound semiconductors [41, 42]. This model states that the persistence photo-conductivity decay is dominated by thermal activation of carriers to overcome the electron-capture barrier in the higher temperature region while at low temperatures the capture occurs by tunneling via multi-phonon processes [42, 43]. A similar argument could be well associated in the present case for the oxide heterostructures. However, it is interesting to note that the photo-response in the LTO/STO and LAO/STO systems shows a significant drop on decreasing the growth temperature, with the response remaining sensitive to UV radiation only. In this context we emphasize that with increasing the growth temperature, defects in the form of O vacancies may be created in STO substrate just below the interface, driving the interfacial Ti ions to exhibit multi-valent states. Isolated oxygen vacancies in STO imply a 3d$^1$ configuration of Ti in its neighborhood, whose scattering cross-section would be very different from that of a Ti ion with 3d$^0$ electronic configuration. Thus defect induced localized states, which are prominent in samples grown at high temperatures are expected to have more traps as compared to the that of samples with abrupt interface grown at low temperatures. This suggests that the decay of UV generated photo-current can be precisely controlled depending on the growth conditions of the sample. The value of $\tau$ at 20 K varies from $\sim$3950 s to 1750 s and from $\sim$2955 s to 1700 s for LTO/STO and LAO/STO respectively as growth temperature is decreased from 800$^o$C to 700$^o$C. On increasing the growth temperature the activation energy increases from $\sim$2 - 11 meV and $\sim$7 - 20 meV in the high temperature regime for LTO and LAO respectively and remains $\sim$1 meV in low temperature region, while the value of exponent $\beta$ is almost unaffected by growth temperature and varies from 0.2 - 0.8 throughout the temperature region.

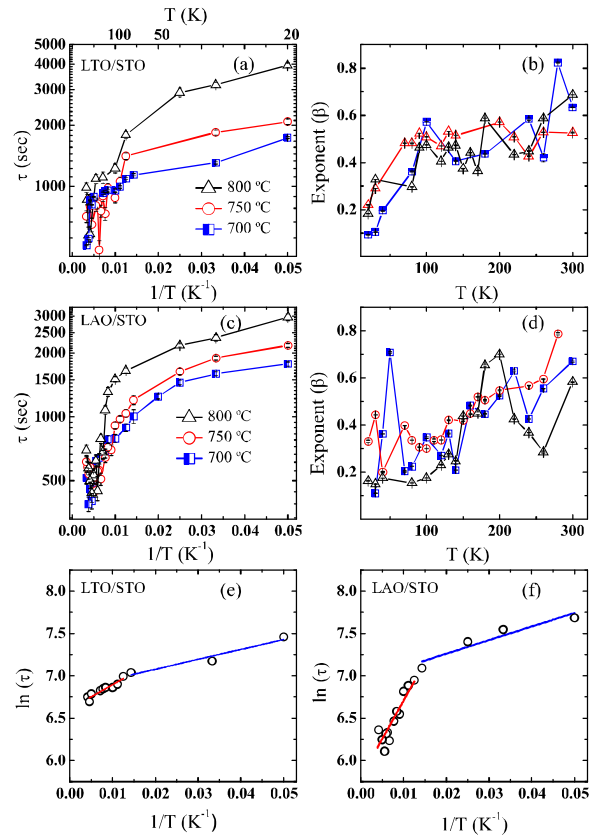

Figure 4: (Color Online) Critical parameters $\tau$ and $\beta$ of the Kohlrausch expression for LTO/STO and LAO/STO systems. The variation of the time constant with 1/T is plotted in semi log scale in panel (a) and (c) for LTO/STO and LAO/STO films respectively, grown at different $T_D$. The dependence of exponent is plotted in panel (b) and (d) respectively. The values of $\tau$ and $\beta$ were extracted from fitting to the photo-response at different temperatures to the Kohlrausch expression. Temperature dependence of time constant can be fitted to the Arrhenius equation with two different activation energies as shown in panels (e) and (f).

The behavior of the photo-conductivity decay in the oxide heterostructures appears quite similar to that seen in $Ga_{0.3}Al_{0.7}As$, [42] $Ga_{0.8}Al_{0.2}N$, [35] and $Zn_{0.04}Cd_{0.96}Te$, [44] systems. The activation energy reported for these systems fall in the range of ∼300 to 100 meV, which is one order higher in magnitude when compared to the LTO/STO and LAO/STO systems (10 - 20 meV). In general, smaller activation energies point to weak lattice relaxation [45]. Here, we note that the major factors that govern the relaxation effects in $Ga_{1-x}Al_xAs$ and oxide heterostructures are different. In $Ga_{1-x}Al_xAs$ and related systems, the local relaxation fundamentally stems from strain effects due to ionic size mismatch [46, 47], while in oxide heterostructures the origin is due to polar catastrophe. Thus, the measure of low activation energies in the LTO/STO and LAO/STO heterostructures suggest that the intrinsic electric field induced redistribution of electronic states, due to the polar discontinuity, across the interface is rather weakly coupled to the lattice, thereby leading to small activation energies. However, the increase in the activation energy for the high temperature grown heterostructures may be attributed to the creation of oxygen vacancies, which couple strongly to the lattice.

### D. Effect of electric field on optical response

It has been shown earlier that the electrical conductivity of these structures can be modulated significantly by applying an electric field[17, 27, 30]. Here we examine the temporal dependence of this effect in order to compare it with the time evolution of the photo-conducting state. Figure 5 (a) and (b) show the effects of gate field $E_g$ of ± 1, 2, and 4 kV/cm on these systems, which was applied in a back gate geometry. The channel resistance drops spontaneously on application of + $E_g$ and remains constant in the time as long as the gate field is on. The system recovers quickly in less than a microsecond time scale on turning off the field. The effect of the negative gate field is to increase the resistance of the channel. However, the response time remains the same as seen for +$E_g$. The current-voltage curves of the channel for both polarity of fields remain ohmic throughout the field range. The percentage change in channel resistance, defined as $\Delta R = \frac{|R(0)-R(E_g)|}{R(0)} \times 100$, for LTO and LAO at $E_g$ = 4 kV/cm is 19.21 and 15.55 respectively. A dramatically different time dependence of $\Delta R$ seen during recovery from the photo-excitation and gating processes strongly suggest that these two fields act on different type of carriers. In order to arrive at a definite conclusion on this issue, we have done experiments in which the photon and electrostatic fields are applied in tandem.

Fig. 6 shows the drop in resistance of LTO and LAO heterostructures on photo exposure at 20 K. The samples were subjected to ± $E_g$ during the recovery after 0, 100, 500,1000, and 1500 seconds of turning off the light. An abrupt change in resistance takes place on application of the gate field. The resistance drops with +$E_g$ whereas rises with -$E_g$. In inset of Fig. 6 we plot the instantaneous change in R on application of ±$E_g$ at several instants of time. The data show that the change in R due to electric field is nearly independent of the state of recovery of the photo-conducting state. This confirms our earlier observation that the two fields act on two different sources of carriers.

The recovery of the system from the photo-conducting state at several values of the gate voltage is shown in Fig. 7(a& b), for LTO and LAO respectively. The gate field was applied at the same time when the light was turned off. We clearly see that the recovery is accelerated by -$E_g$ whereas +$E_g$ slows it down. In the case of LAO/STO we also note that the negative field actually pushes the resistance beyond the value in dark. in Fig. 7(c) we plot the change in resistance $\Delta R$ after 1000 seconds of the applying the gate field. Here $\Delta R$ is defined as $\Delta R = \frac{|R(1000)_{E_g=0} - R(1000)_{E_g})|}{R(1000)_{E_g=0}}$

### E. Electronic structure and optical conductivity

To address the wavelength dependence of the photo-current and to explain the decay of the photo-excited

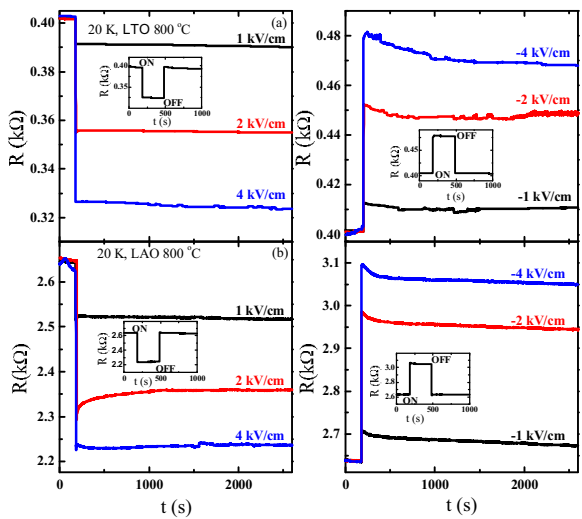

Figure 5: (Color Online) Panel (a) and (b) show the effect of ± gate field at 20 K on the LTO/STO and LAO/STO thin films fabricated in three terminal configuration. After switching on the field the resistance is measured as a function of time. The effect of gate field on the conductivity of the channel is much faster as compare to the effect of photo-excitation. Positive field makes channel conducting while negative makes it more resistive. The effect of successive on and off cycles is shown in the inset.

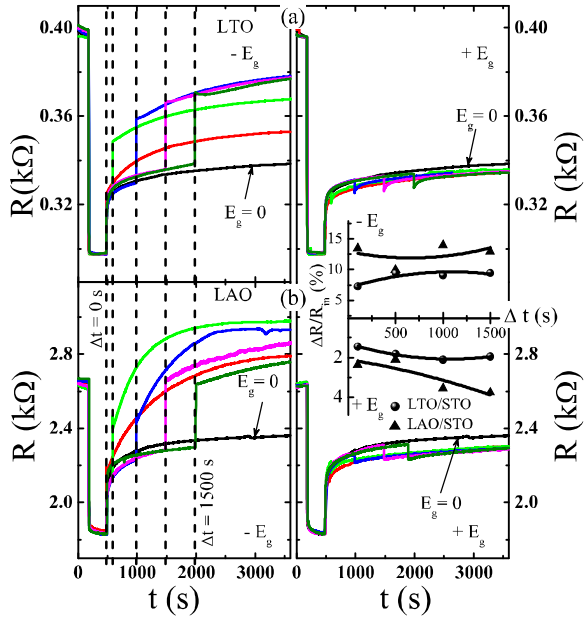

Figure 6: (Color Online) The combine effect of photo exposure and gate field ±2kV/cm. on (a) LTO/STO and (b) LAO/STO is examined at 20 K. The field is applied after certain time intervals (dotted lines, $\Delta t$ = 0, 100, 500, 1000 and 1500 s) during the recovery of the system after extinguishing the photo exposure. The recovery of the resistance with the applied gate field is shown for the sake of comparison. It is clear that the -ve field increases abruptly and significantly, whereas the effect of +ve field is small but still abrupt. The inset shows the relative change in the resistance ($\Delta R/R_{in}$) at the time when the gate field is applied. Here $\Delta R = |R_{in} - R_{final}|$ and X- axis is the time at which the field was turned on after terminating the photo exposure.

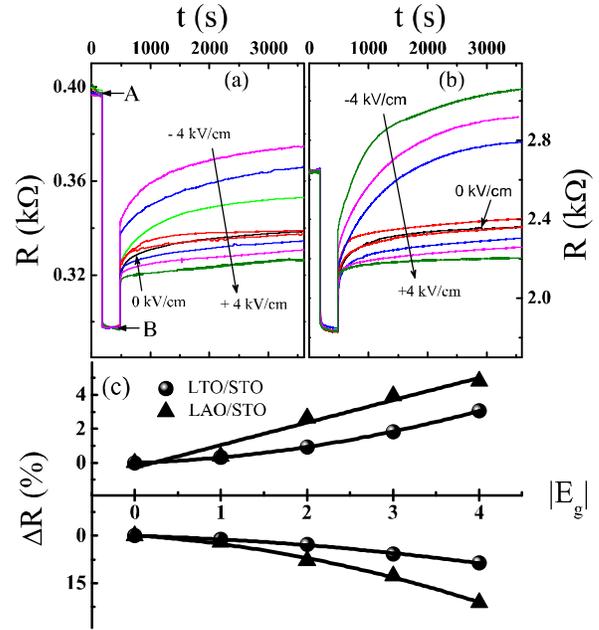

Figure 7: (Color Online) The recovery to the initial state under applied gate fields immediately after terminating the light exposure. The panel (a) & (b) are for LTO/STO and LAO/STO respectively grown at 800 $^oC$. The change in resistance ($\Delta R$) at t = 1000 s as a function of gate field is shown in (c). At point 'A', shown in the panel (a), the photo exposure with ∼ 8$\mu$W/cm$^2$ light was irradiated, while at point 'B' the exposure is stopped and the system recovery is monitored as a function of time under different gate fields applied at instant 'B'. It can be clearly seen that the fields of both polarities alter the recovery process; the conducting state of the system is arrested when + $E_g$ is applied and the recovery higher resistibly is accelerated with - $E_g$.

carriers, we study the band structure of the structurally relaxed LAO/STO system.

With the GGA formalism, the energy gap of the TiO$_2$ terminated SrTiO$_3$ was found to be 0.81 eV, while with two and three LAO layers the energy gap was reduced to 0.71 eV and 0.02 eV, respectively. For four unit-cells of LAO on TiO$_2$ terminated STO substrate the ground state was found metallic.

In Fig.8, we show the ground state of the fully relaxed 2×2×9 dimension super-cell with four units of LAO on TiO$_2$ terminated STO. In comparison with the semiconductor heterostructures such as AlGaAs/GaAs systems, which also consists of two electronically gapped compounds sharing a common anion element, the valence band wave functions in both LAO-STO and AlGaAs/GaAs heterostructures is seen to evolve mainly from the atomic wave function of anions, while the conduction band wave functions evolve mainly from the atomic wave functions of cations. In case of LAO/STO systems, there are few bands crossing the Fermi energy ($E_F$). The metallicity is due to the overlap of the surface O 2p bands of the AlO$_2$ layer with that of the Ti 3d bands of the TiO$_2$ interface, which is consistent with the previous reports [48]. Furthermore, analysis of the band character shows that the states below $E_F$ emanate mainly from O 2p orbitals. The top of the valence bands are primarily composed of O 2p states resulting from the AlO$_2$ layers of the LAO film. With increasing energy and for E - $E_F$ > 2 eV, we find that the states that constitute the band structure are essentially O 2p states of the TiO$_2$ layers of the STO motif. In other words, we find that the valence band of the LAO-STO system depicts a hierarchy in the distribution of O 2p states as a function of increasing binding energy with widely distributed O 2p states of AlO$_2$ layers forming the upper valence states ($E_F$<E <-2 eV), and that from the TiO$_2$ layers forming the intermediate valence band which are relatively more localized over the energy range -2 eV<E<-3 eV.

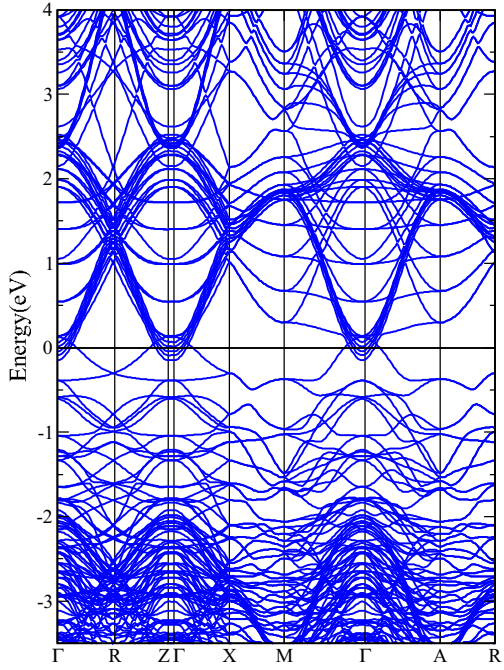

Figure 8: The FP-LAPW generated band structure of the fully relaxed LaAlO$_3$-SrTiO$_3$ heterostructure. The zero in the energy scale represents the Fermi energy.

On the other hand, the states above E$_F$ are primarily composed of Ti 3d orbitals. Orbital decomposition of the states that constitute the bands crossing at E$_F$ reveals that they are the t$_{2g}$ states of Ti 3d orbitals of the TiO$_2$ interface, which are strongly hybridized with O 2p states of the surface AlO$_2$ layer. In the energy range E$_F$ < E < 2 eV, band structure traces a series of parallel sub-bands along the Brillouin zone edges. These are the 3d$_{xy}$ bands of the Ti ions of the STO motif. The lowest of these parallel sub-bands emerges from the interfacial TiO$_2$ layer. With increasing energy in the unoccupied region of the band structure, these parallel sub-bands are found to emerge in succession from the 3d states of the TiO$_2$ layers of the STO motif, away from the interface. At E ≈ 2 eV above E$_F$, one finds a conglomeration of Ti 3d bands of e$_g$ character, localized over a very narrow energy interval. This localization of the e$_g$ states is quite evident in the density of states spectrum as shown in Fig. 9(a).

From the band structure, we see that the number of bands crossing the E$_F$ are small, while there is a bunching of bands few electron-volts (≈ 2 eV) above and below E$_F$. Therefore, one may expect that photo-conductivity (PC), which results from inter-band transitions, would occur between these states and the states at E$_F$, i.e., corresponding to the wavelength less than 500 nm. Experimentally, this is observed for λ < 440 nm. Thus, in order to substantiate our arguments, we have calculated the conductivity σ(ω) using the Eq. 1 and Eq. 2 from Ref. [49], which for the imaginary part of the dielectric function arising from inter-band transitions, is given as,

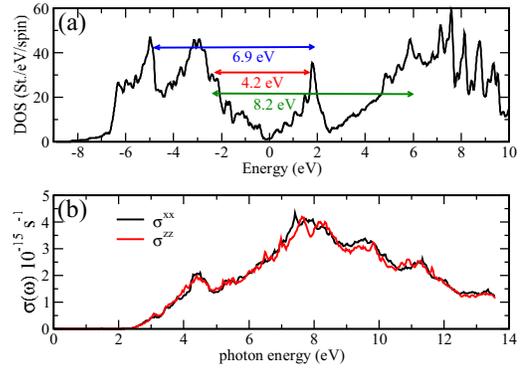

Figure 9: (Color Online) The DFT generated total density of states (DOS), of the fully relaxed LAO/STO heterostructure showing the most probable transitions that can take place between the valence and conduction bands, and (b) optical conductivity ($\sigma(\omega)$) for the two polarizations corresponding to the electric field in the basal plane ($xx$ component) and along the c-axis ($zz$ component). For E < 2.5 eV, $\sigma(\omega)$ is small. It becomes significantly larger for E > 2.5 eV (λ≈ 500 nm), which is consistent with the experimental data where only a small intensity is observed for λ > 440 nm. Note that there is a reasonably good agreement between the transition energies shown in (a) and the $\sigma(\omega)$ peak positions shown in (b).

$$\epsilon_2^\alpha(\omega) = \frac{4\pi^2 e^2}{m^2\omega^2} \sum_{v,c} \frac{2}{V} \sum |\langle\psi_{c,k}|p_\alpha|\psi_{v,k}\rangle|^2 \delta[E_c(k) - E_v(k) - \hbar\omega] \quad (1)$$

where α=(x, y, z) is the axis index of tetragonal cell, $v$ and $c$ represents the valence and conduction band respectively and V the super-cell volume including the vacuum region. The optical conductivity σ(ω) is given as

$$\sigma(\omega) = \frac{\omega\epsilon_2(\omega)}{4\pi} \quad (2)$$

The calculated σ(ω) is shown in Fig.9 (b). We find that for E < 2.5 eV, σ(ω) is small, and that it becomes significantly larger for E > 2.5 eV (λ≈ 500 nm). This is consistent with the experimental data where only a small photo-conductivity is observed for λ > 440 nm. Also, one may note that the mean field nature of density functional theory limits the accuracy of band positions, and thus we expect certain degree of uncertainty in the location of peaks in the calculated optical properties. Our calculations also find that there exists a number of peaks in σ(ω) located at around 4.4, 7.5, 9.5 and 11.3 eV. Peaks in σ(ω) arise from transitions involving parallel bands. These can be identified from the band structure (Fig.

8) or from the density of states (Fig. 9(a)). We have shown in Fig. 9(a) the possible transitions which could lead to the 4.4 and 7.5 eV peaks in $\sigma(\omega)$. We mention that this is only indicative as the correct structure is obtained from the calculations based on Eqs.1-2. We have calculated $\sigma(\omega)$ for the two polarizations corresponding to the electric field in the basal plane ($xx$ component) and along the c-axis ($zz$ component). The anisotropy is very small reminiscent of the pristine cubic structures of LAO and STO. There is a reasonably good agreement between the transition energies shown in Fig. 9(a) and the peak positions as shown in Fig. 9(b).

For both LAO/STO and LTO/STO heterostructures, the recovery from the UV radiation field photoconducting state follows the stretched exponential decay. The overall nature of the decay seems to be independent of the film growth conditions on the $TiO_2$ terminated STO. For low carrier density systems, such as for films grown at lower temperatures, the interface is atomically abrupt, thus being devoid of vacancy induced states to act as traps. Thus, we argue that the source of the such trapping centers in oxide heterostructures are the closely spaced Ti $3d_{xy}$ bands. As seen from the band diagram of Fig. 8 for LAO/STO systems, the Ti $3d_{xy}$ states form flat parallel bands above the $E_F$. Thus, according to the uncertainty principle, $\Delta\epsilon \times \Delta t \geq \hbar$, where $\Delta\epsilon$ represents the band-width and $\Delta t$ the carrier life-time, it may be argued that for highly localized Ti $3d_{xy}$ states, the electrons trapped in these states would have longer life-time, resulting in a slower decay.

However, in comparison with the LAO-STO, the LTO-STO systems show a relatively larger decay constant. This may be partly attributed to the strong oxidation tendency of the $LaTiO_3$ by imbibing oxygen from the interfacial layers of the STO substrate. It has been reported before that the O vacancies in dilute concentrations would create new states in the band gap of $SrTiO_3$ below the conduction band [50–52], representing a shallow state. These defect induced states are also localized and may act as additional traps to the photo-excited carriers. Thus, one would expect that the LTO-STO and also those systems which have been synthesized at higher temperatures that carry vacancies would have higher decay rate.

The sudden decrease in the photo-conductivity on application of a gate voltage can also be reasonably well accounted on the basis of the band picture. The excited states (conduction band states) of these oxide heterostructures are derived from the Ti 3d states of the STO motif, which is a incipient ferroelectric. It has been previously shown that $SrTiO_3$ display ferroelectric characteristics at low temperatures as well when subjected to an strain and external electric field [53–55]. Following the $d^0$-ness model, which explains ferroelectric properties of the iso-electronic $BaTiO_3$ systems [56, 57], one may then expect an off-centering of the Ti ions in the first few layers of substrate STO. Under these circumstances, changes in the Ti-O hybridization can therefore lead to dispersive bands. We expect that on application of electric field the parallel Ti $3d_{xy}$ sub-bands which act as trapping centers due to their localized nature, would be broadened and therefore decrease the carrier life-time in these bands.

## IV. SUMMARY AND CONCLUSIONS

In summary, we have performed a comparative study electronic transport of LTO/STO and LAO/STO heterostructures under the influence of radiation and electric fields, applied both separately and simultaneously . The heterostructures are found sensitive to the near ultraviolet radiation which is well supported by the optical calculations based on band theory. Both the systems show persistence photo-conductivity, which follows a stretched exponential behavior. The difference in the energy scale for recovery in two temperature regions suggests a change in the migration dynamics of photo-excited carriers with thermal energy available to the system. It is also pointed out that the change in resistance due to electric field is almost independent of the state of recovery of the photoconducting state which confirms that the two fields act on two different sets of carriers. The observed photoresponse is large for LTO/STO and is greatly affected by the growth temperature, i.e., higher for the films grown at higher temperatures, thus eluding towards a role of oxygen vacancies created in STO during the film deposition. We argue that these localized vacancy states act like additional trap states for the carriers thereby inducing the slow relaxation. However, the primary factor that governs the photo-current decay is the Ti derived $3d_{xy}$ states, which in the lower conduction band unveil as highly localized states. Following the uncertainty principle, localization of states lead to increased carrier life-times, and thus for carriers trapped in these Ti $3d_{xy}$ states are expected to have longer life-time, resulting in a slower decay. Thus, the relaxation dynamics of the charge carriers seems to be associated to the intrinsic property of the $SrTiO_3$ substrates in these heterostructures.Furthermore, the dependence of sheet charge density with the photon flux and wavelength can bring new understanding to these system.

### Acknowledgments

AR would like to acknowledge the Council of Scientific and Industrial Research (CSIR), India and Indian Institute of Technology Kanpur for financial support. RCB acknowledges the J. C. Bose National Fellowship of the Department of Science and Technology, Government of India.